\documentclass[aps,prc,12pt,nofootinbib,superscriptaddress,preprintnumbers]{revtex4-1}
\usepackage{selinput}
\usepackage[T1]{fontenc}
\usepackage{graphicx}
\usepackage{epstopdf}
\usepackage{amsmath,amssymb}
\usepackage{verbatim}
\usepackage[dvipsnames]{xcolor}
\usepackage{geometry}
\usepackage{units}
\usepackage{systeme}
\usepackage{amsmath}
\usepackage[toc,page]{appendix}\usepackage{float}
\usepackage{subcaption}
\usepackage{lipsum}
\usepackage{hyperref}
\usepackage{caption}
\usepackage[normalem]{ulem}
\begin{document}
\title{Quark saturation in the QCD phase diagram}

\author{Marcus Bluhm}
\affiliation{SUBATECH UMR 6457 (IMT Atlantique, Universit\'e de Nantes, IN2P3/CNRS)\\ 4 rue Alfred Kastler, 44307 Nantes, France}
\author{Yuki Fujimoto}
\affiliation{Institute for Nuclear Theory, University of Washington\\
  Box 351550, Seattle, WA 98195, USA}
\author{Larry McLerran}
\affiliation{Institute for Nuclear Theory, University of Washington\\
  Box 351550, Seattle, WA 98195, USA}
\author{Marlene Nahrgang}
\affiliation{SUBATECH UMR 6457 (IMT Atlantique, Universit\'e de Nantes, IN2P3/CNRS)\\ 4 rue Alfred Kastler, 44307 Nantes, France}

\date{\today}

\preprint{INT-PUB-24-048}

\newcommand{\MN}[1]{\textcolor{magenta}{#1}}
\newcommand{\MB}[1]{\textcolor{MidnightBlue}{#1}}
\newcommand{\YF}[1]{\textcolor{Orange}{#1}}
\newcommand{\LM}[1]{\textcolor{Red}{#1}}

\newcommand{\kFB}{k_{\rm FB}}
\newcommand{\EF}{E_{\rm F}}
\newcommand{\kFQ}{k_{\rm FQ}}
\newcommand{\kbu}{{k_{\rm bu}}}
\newcommand{\ksh}{{k_{\rm sh}}}
\newcommand{\qbu}{{q_{\rm bu}}}
\newcommand{\qsh}{{q_{\rm sh}}}
\newcommand{\Nc}{N_{\rm c}}
\newcommand{\DB}{\Delta_{\rm B}}
\newcommand{\DQ}{\Delta_{\rm Q}}
\newcommand{\fFD}{f_{\rm FD}}
\newcommand{\fB}{f_{\rm B}}
\newcommand{\betaid}{\beta_{\rm id}}
\newcommand{\muid}{\mu_{\rm id}}
\newcommand{\Tid}{T_{\rm id}}

\newcommand{\kF}{k_{F}}
\newcommand{\muB}{\mu_{B}}
\newcommand{\muBsat}{\mu_{B,\mathrm{sat}}}

\begin{abstract}
We determine the onset of Quarkyonic Matter corresponding to values of temperature and baryon chemical potential at which the quark phase space density becomes one. At zero temperature for baryon chemical potentials below the mass of the Lambda baryon, only nucleons contribute to the quark density. This is different at finite temperature, where all baryons, mesons and their resonances can be excited and thus add quarks to the phase space. The probability density to find a quark inside a hadron is determined using the Yukawa ansatz of the IdylliQ model of Quarkyonic Matter. We estimate separately the magnitude of the various contributions of nucleons, Delta baryons, pions as well as further hadrons and resonances. The uncertainty in the parametrization of the probability density to find a quark inside a nucleon is spanned by assuming that at zero temperature the transition density to Quarkyonic Matter is between one and three times that of nuclear matter. Various predictions for a possible critical point associated with the chiral phase transition are found close to a triple point at which the line of the deconfinement transition and the curve associated with the transition to Quarkyonic Matter intersect. These considerations provide an estimate for the region in the QCD phase diagram where Quarkyonic Matter may be found.
\end{abstract}

\maketitle

\section{Introduction}

It has been conjectured that there is a transition of nuclear matter to Quarkyonic Matter~\cite{McLerran:2007qj} at high baryon density. At zero temperature, this matter exists at some intermediate baryon chemical potential $\muB$ between a typical QCD scale, $\muB \sim M_N$, and the scale at which deconfinement occurs, $\muB \sim \sqrt{N_c} M_N$. Here, $M_N$ is the nucleon mass and $N_c$ is the number of quark colors, which for our world is $N_c = 3$. While the arguments presented in Ref.~\cite{McLerran:2007qj} were for a large number of colors, in Ref.~\cite{McLerran:2018hbz} it was argued that for $N_c = 3$ such an intermediate baryon density phase of Quarkyonic Matter has the correct semi-quantitative properties to explain the very stiff equation of state needed to describe recent neutron star data (see, e.g., Refs.~\cite{Bedaque:2014sqa, Tews:2018kmu, Fujimoto:2019hxv, *Fujimoto:2021zas, *Fujimoto:2024cyv, Drischler:2020fvz}).

At $T=0$, this ability to explain the rapid stiffening of the equation of state within Quarkyonic Matter, as was also shown in~\cite{Fujimoto:2023mzy}, arises as follows: once the phase space density of quarks becomes one, the momentum distribution of nucleons is of occupation number one in a momentum space shell near the Fermi surface. Below, in the Fermi sea, the nucleon distribution is under-occupied with occupation number of order $1/N_c^3$. Because of the under-occupied phase space of nucleons in the sea, the nucleons may become relativistic without causing the baryon density to increase too rapidly, a property needed for a stiff equation of state. This is because the Fermi momentum increases rapidly as a function of baryon density for an under-occupied distribution. The quarks that correspond to this nucleon distribution are determined by the probability of finding a quark inside a nucleon convoluted with the nucleon distribution. Their distribution corresponds to a filled Fermi sea of quarks up to a bulk momentum $q_{\rm bu} \le k_{\rm bu}/N_c$. Here, $q_{\rm bu}$ is a Fermi momentum that corresponds to the under-occupied part of the nucleon distribution at $T=0$. There is an exponentially falling distribution for quarks at the quark Fermi surface, which arises from the effect of confinement, corresponding to the fully occupied distribution of nucleons at the Fermi surface. To describe the stiff equation of state required for neutron stars, the transition to Quarkyonic Matter must occur at a density of the order of that of nuclear matter.
Indeed, it was pointed out in Ref.~\cite{Drischler:2020fvz} that the small tidal deformability from GW170817~\cite{LIGOScientific:2017vwq} and the calculations of the equation of state from the chiral effective field theory~\cite{Drischler:2020hwi} together imply that the rapid stiffening should occur around $\gtrsim 1.5\mbox{--}1.8$ times the nuclear matter density.

Surprisingly, the estimation of the baryon density necessary for the onset of Quarkyonic Matter does not require the knowledge of its dynamic properties. Such an estimate needs only the computation of the phase space density $f_Q$ of quarks inside of nucleons in ordinary nuclear matter. The baryon density at which Quarkyonic Matter occurs is the density at which the phase space density of quarks at some momentum saturates and becomes $1$. It is reasonable to assume that this first happens at zero quark momentum, $q=0$. Simple estimates for zero temperature baryonic matter can be made using experimentally constrained form factors or unintegrated valence and sea quark distribution functions~\cite{Koch:2024zag,McLerran:2024rvk}. Such estimates suggest that the onset of Quarkyonic Matter occurs at a density quite close to that of nuclear matter. In this paper, we will take a range larger than is predicted from these computations, with a density in the range of 1--3 times that of nuclear matter.

At finite temperature, the computation of the onset density of Quarkyonic Matter may be carried out along the same lines (see also~\cite{Kojo:2022psi}). This is fortunate as there is not yet consensus about how to write down a proper description of Quarkyonic Matter at finite temperature~\cite{Sen:2020peq}. In this paper, we present the determination of the onset of quark saturation in the QCD phase diagram. For this, we generalize the ideal gas of hadrons that was the basis of the Ideal Dual Quarkyonic (IdylliQ) model for nucleons, to a resonance gas of non-interacting hadrons. The contributions to the phase space density of quarks arise from the baryons to leading order in $N_c$ and a $1/N_c^2$ correction from the mesons. In this paper, we compute the values of temperature and baryon chemical potential corresponding to the transition density at which quark saturation occurs. In later papers, we will turn to the problem of the proper formulation of an IdylliQ model and a quantum statistics formulation for Quarkyonic Matter at finite temperature.

By going from zero to finite temperature, we realize that the Quarkyonic Matter description is no longer valid in the deconfined phase and therefore has a natural upper temperature limit of applicability. We present a variety of arguments to determine this boundary of our computation. One is simply the large $N_c$ limit, where the deconfinement temperature is independent of $\muB$. This can be corrected by computing the region where the Debye screening mass equals the QCD scale $\Lambda$. This is also compared with what is known at low baryon density from lattice gauge theory. Finally, we can compare the boundary formed by the Quarkyonic Matter region with that of the deconfinement transition to the range of decoupling temperatures supposedly measured in heavy-ion collisions. 

Our result for the boundary region for new physics is quantitatively similar to that presented and discussed in~\cite{Andronic:2009gj}. We defer to this paper a discussion of the relationship between the Quarkyonic Matter description and other attempts to characterize the region where there is new physics. We believe the determination we present in this paper provides a stronger theoretical and quantitative basis than these earlier computations. We also compare our results to several recent estimates for the position of the critical end point, where a first-order chiral transition is supposed to terminate in a second-order critical point. We find that the triple point where the transitions between hadronic matter, deconfined matter, and Quarkyonic Matter all meet, is close to where this critical endpoint is estimated to be found.

\section{Quark saturation in a nucleon gas at finite temperature\label{sec:2}}

Coming from the dilute, ordinary nuclear matter regime and increasing the density, we can find quark\footnote{We would like to underline that we are only looking at quarks and not anti-quarks as the anti-quarks are highly suppressed at finite $\muB$. Thus all densities in this work refer to the quark content of particles and are not net-densities.} saturation from the ideal gas thermodynamics under the condition that $f_Q(q=0)=1$.
For a given color, the quark distribution function per quark degree of freedom can be obtained from the nucleon distribution function via~\cite{Fujimoto:2023mzy}
\begin{equation}
   f_Q(q) = \int \frac{d^3k}{(2\pi)^3}\, \varphi(|\vec{q}-\vec{k}/N_c|)\, f^N_{\rm FD}(k) \,,
   \label{eq:fQ}
\end{equation}
where $\varphi(p)$ denotes the momentum distribution of quarks inside the nucleons and 
$f^N_{\rm FD}(k)$ is the Fermi-Dirac distribution function of nucleons per nucleon degree of freedom reading 
\begin{equation}
    f^N_{\rm FD}(k) = \frac{1}{1+e^{(E_N(k)-\mu_N)/T}} \,.
    \label{eq:fN}
\end{equation}
In Eq.~\eqref{eq:fN}, $E_N(k)=\sqrt{k^2+M_N^2}$ is the energy of the nucleons with mass $M_N$, $T$ is the temperature and $\mu_N$ is the nucleon chemical potential. Assuming $\mu_Q=\mu_S=0$ for electric charge and strangeness chemical potentials in the following, we have $\mu_N=\muB$. 

The condition for quark saturation, which marks the onset of Quarkyonic Matter formation at high density, then reads
\begin{equation}
 1 = f_Q(q=0) =  \frac{d_N}{d_Q} \int \frac{d^3k}{(2\pi)^3}\,\varphi(k/N_c) f^N_{\rm FD}(k)\,,
 \label{eq:cond_onset}
\end{equation}
where in Eq.~\eqref{eq:cond_onset} we take the number of isospin-spin degrees of freedom of quarks and nucleons to be $d_Q=d_N=4$. In~\cite{Fujimoto:2023mzy}, the momentum distribution of quarks inside a nucleon was assumed to be 
\begin{equation}
  \varphi(p) = \frac{2\pi^2}{\Lambda^3} \frac{\exp({-|\vec{p}\,|}/\Lambda)}{|\vec{p}\,|/\Lambda}\,.
\label{eq:phiq}
\end{equation}
Here, the QCD scale $\Lambda$ gives the relevant wavefunction momentum. This is the ansatz of the IdylliQ model which we will follow also in this work. At $T=0$ we can evaluate the condition Eq.~\eqref{eq:cond_onset} for this choice of $\varphi(p)$ analytically and find 
\begin{equation}
    1 = N_c^3 \left[1-e^{-\kF/(N_c \Lambda)}\left(1+\frac{\kF}{N_c \Lambda}\right)\right]\,,
    \label{eq:cond_onsetT=0}
\end{equation}
where $\kF=\sqrt{\muB^2-M_N^2}$ is the Fermi momentum at $T=0$. The saturation takes place at low $\kF$, i.e.~$\kF\ll N_c \Lambda$. With this approximation, one obtains $\kF=\Lambda\sqrt{2/N_c}$ from Eq.~\eqref{eq:cond_onsetT=0} as a condition for the onset of quark saturation. This condition had already been determined in~\cite{Fujimoto:2023mzy}. The corresponding baryon chemical potential for the onset of saturation is $\muBsat(T=0)=\muB^*=\sqrt{2\Lambda^2/N_c+M_N^2}$.

The behavior of $\muBsat(T)$ at finite temperature can be estimated analytically.
Because the low momentum part of the Fermi-Dirac distribution function dominantly contributes to the quark saturation, one can approximate $E_N(k) \simeq k^2/(2M_N) + M_N$ and $\varphi(k/N_c) \simeq 2 \pi^2 N_c /(\Lambda^2 k)$. With those approximations, Eq.~\eqref{eq:cond_onset} can be evaluated analytically to give 
\begin{equation}
    \muBsat(T) \simeq T \ln \left[e^{\Lambda^2 / (N_c M_N T)} - 1\right] + M_N\,.
\end{equation}
At low temperature, $T \lesssim \Lambda^2 / (N_c M_N) \sim \Lambda / N_c^2$, the saturation curve does not depend on $T$,
\begin{equation}
    \muBsat(T) \simeq \frac{\Lambda^2}{N_c M_N} + M_N\,.
    \label{eq:SF-SaturationCondition_AltLow}
\end{equation}
At high temperature, $T \gtrsim \Lambda^2 / (N_c M_N)$, one obtains $\muBsat(T)$ as a decreasing function of $T$, and thus the saturation curve bends towards smaller $\mu_B$ in the $T$-$\muB$ plane with increasing $T$ as 
\begin{equation}
    \muBsat(T) \simeq - T \ln \left[\frac{T}{\Lambda^2 / (N_c M_N)}\right] + M_N\,.
    \label{eq:SF-SaturationCondition_Alt}
\end{equation}

At small temperatures, one may also evaluate the integral in Eq.~\eqref{eq:cond_onset} by means of a Sommerfeld expansion, see Appendix~\ref{sec:A1}. In the limit of large $N_c$, where we can approximate $\varphi(k/N_c) \simeq 2 \pi^2 N_c /(\Lambda^2 k)$, this amounts to 
\begin{equation}
    \muBsat(T) = \sqrt{2\left(\frac{\Lambda^2}{\Nc} - \frac{\pi^2}{6}\, T^2\right)+M_N^2}
    \label{eq:SF-SaturationCondition}
\end{equation}
for the onset baryon chemical potential as a function of $T$. 
According to Eq.~\eqref{eq:SF-SaturationCondition}, one also finds $\muBsat(T)$ as a decreasing function of $T$. 
In the $T-\muB$ phase diagram this results in a curve depicting the transition to Quarkyonic Matter similar to the liquid-gas phase transition, with 
\begin{equation}
    T_{\rm sat}(\muB) = \sqrt{\frac{6}{\pi^2}\left(\frac{\Lambda^2}{\Nc}-\frac{1}{2}\left(\muB^2 - M_N^2\right)\right)}\,.
\end{equation}
We note that the two provided estimates Eqs.~\eqref{eq:SF-SaturationCondition_AltLow} and~\eqref{eq:SF-SaturationCondition} coincide in the large $N_c$ limit when $T\sim\Lambda^2/(N_c M_N)$. 

\begin{figure}[t]
      \centering
      \includegraphics[width=0.75\textwidth]{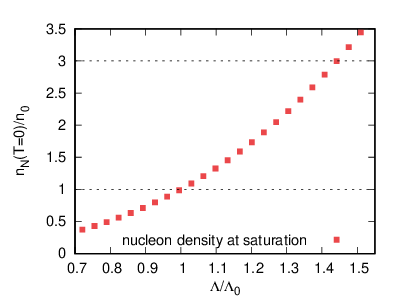}
      \caption{Scaled nucleon density at quark saturation for $T=0$ as a function of $\Lambda$ in units of $\Lambda_0\simeq 0.29$~GeV with $N_c=3$ and nuclear matter density $n_0=0.16~$fm$^{-3}$.}
      \label{fig:nNdensSatT0}
\end{figure}
We may study the onset condition of quark saturation as a function of wavefunction momentum $\Lambda$. For this reason, we determine the nucleon density at $T=0$ via 
\begin{equation}
    \label{eq:nNSatT0}
    n_N(T=0) = \frac{d_N}{2\pi^2} \int_{M_N}^{\muB^*} dE_N\,E_N\sqrt{E_N^2-M_N^2} 
    = \frac{d_N}{6\pi^2}\left((\muB^*)^2-M_N^2\right)^{3/2}\,.
\end{equation}
The corresponding result is shown in Fig.~\ref{fig:nNdensSatT0}. To be consistent with neutron star phenomenology, we expect the nucleon density for Quarkyonic Matter formation to lie in between nuclear matter density $n_0$ and about three times nuclear matter density. This gives us an uncertainty band in $\Lambda$ between $1\leq \Lambda/\Lambda_0 \leq 1.44$, see Fig.~\ref{fig:nNdensSatT0}, which we will explore below. Here, $\Lambda_0\simeq 0.29$~GeV for which $n_N(T=0)=n_0$.

To discuss the behavior of the nucleon density for $T>0$, a Sommerfeld expansion similar to the one outlined in Appendix~\ref{sec:A1} can be done. From this we find to leading order in $T$
\begin{align}
    \nonumber
    n_N(T,\muB) & = \frac{d_N}{2\pi^2} \int_0^\infty dk \,k^2 f^N_{\rm FD}(k) \\
    \label{eq:nNdensSommerfeld}
    & \simeq \frac{d_N}{6\pi^2} \left(\muB^2-M_N^2\right)^{3/2} + 
    \frac{d_N}{12} T^2 \frac{(2\muB^2-M_N^2)}{\sqrt{\muB^2-M_N^2}} \,.
\end{align}
The nucleon density at quark saturation is then obtained by inserting the onset baryon chemical potential in Eq.~\eqref{eq:SF-SaturationCondition} into Eq.~\eqref{eq:nNdensSommerfeld}. This gives a nucleon density which is monotonically increasing with small, increasing $T$ along the saturation curve as
\begin{equation}
    \label{eq:nNdensSatT}
    n_{N,{\rm sat}}(T) \simeq n_N(T=0) + \frac{d_N}{12}\frac{(\muB^*)^2}{\sqrt{(\muB^*)^2-M_N^2}}T^2 \,.
\end{equation}

\section{Quark saturation in a hadron resonance gas}

At finite temperature, the first excitations to the nucleon gas are the pions described as a superposition of color-anticolor states 
\begin{equation}\label{eq:pionwf}
    |\pi\rangle = \frac{1}{\sqrt{\Nc}}\sum_{i=1}^{\Nc} |c_i\bar{c}_i\rangle \,.
\end{equation}
The quarks contained in the pions also contribute to the saturation and the formation of Quarkyonic Matter. For a gas composed of nucleons and pions, there is now an additional contribution to the quark distribution function from Eq.~\eqref{eq:fQ} 
\begin{align}
   f_Q(q)  = & \,\frac{d_N}{d_Q} \int\frac{d^3k}{(2\pi)^3}\,\varphi(|\vec{q}-\vec{k}/N_c|) \,f^N_{\rm FD}(k) \nonumber\\
             & + \frac{1}{\Nc}\frac{d_\pi}{d_Q}\int\frac{d^3k}{(2\pi)^3}\,\varphi(|\vec{q}-\vec{k}/2|) \,f^{\pi}_{\rm BE}(k)\,,
   \label{eq:fQNpi}
\end{align}
where $f^{\pi}_{\rm BE}(k)$ is the Bose-Einstein distribution function for pions given as 
\begin{equation}
    \label{eq:fPi}
    f^{\pi}_{\rm BE}(k) = \frac{1}{e^{(E_{\pi}(k)-\mu_{\pi})/T}-1} \,,
\end{equation}
$d_\pi$ is the pion degeneracy factor and $\mu_{\pi}$ the pion chemical potential. Since $\mu_Q=\mu_S=0$, we have $\mu_{\pi}=0$. From the normalization of the pion wave function in Eq. (\ref{eq:pionwf}), there is a $1/\Nc$ suppression factor in front of the associated contribution to the quark distribution function. Furthermore, we assume that the momentum of the quark inside a pion of momentum $k$ is distributed around approximately $k/2$ (and roughly the same momentum is carried by the anti-quark) in a similar way as the quark momentum is distributed in a nucleon. We may therefore consider the same functional form for the momentum distribution of quarks $\varphi(p)$.

The quark saturation criterion for a nucleon and pion gas now becomes 
\begin{align}
   1  = & \,\frac{d_N}{d_Q}\frac{N_c}{\Lambda^2} \int_0^\infty dk\,k\,e^{-k/(N_c \Lambda)} f^N_{\rm FD}(k) + \frac{d_\pi}{d_Q}\frac{2}{\Nc\Lambda^2}\int_0^\infty dk\,k\,e^{-k/(2 \Lambda)}  f^{\pi}_{\rm BE}(k)
   \nonumber\\
      = & \, \, C_{\rm N}(T,\muB) + C_{\pi}(T)\, .
   \label{eq:fQsatBM}
\end{align}
The mesonic contribution $C_{\pi}$ from the pions is parametrically suppressed compared to the baryonic contribution $C_{\rm N}$ from the nucleons as $C_{\pi}/C_{\rm N} \sim 1/\Nc^2$. 
Moreover, $C_{\pi}$ is solely a function of $T$ when considering $\mu_Q=\mu_S=0$. In principle, we can fix $\mu_Q$ and $\mu_S$ by conservation constraints on electric charge and strangeness in heavy-ion collisions, but this is beyond the scope of this work.

\begin{figure}[t]
      \centering
      \includegraphics[width=0.75\textwidth]{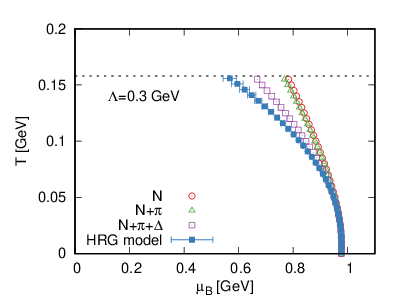}\\
      \caption{(Color online) Saturation transition in the $T$-$\muB$ diagram for $N_c=3$ and $\Lambda=0.3$~GeV considering systems composed of different hadron species. The dashed horizontal line depicts $T_c=0.158$~GeV, the deconfinement transition temperature for $\muB=0$ known from lattice QCD~\cite{Borsanyi:2020fev}. The transition curves are determined for a gas composed of nucleons only (open red circles), nucleons and pions (open green triangles), nucleons, pions and Delta ($\Delta$) baryons (open purple squares) and for a hadron resonance gas (full blue squares) with upper mass limit of $2.25$~GeV. The error bars on the HRG model results are obtained by applying an upper mass cut of $1.6$~GeV, which leads to larger values for $\muB(T)$, and by assuming the same error coming from the impact of unaccounted resonances which are heavier than $2.25$~GeV.}
      \label{fig:HRGSaturation}
\end{figure}

Using the already discussed Sommerfeld expansion for small $T$, one obtains from Eq.~\eqref{eq:fQsatBM} for the nucleon and pion gas in the limit of large $N_c$ 
\begin{equation}
   1  =  \frac{d_N}{d_Q} \frac{N_c}{\Lambda^2} \left(\frac{1}{2}(\muBsat(T)^2 - M_N^2) + \frac{\pi^2}{6}\, T^2\right) + C_{\pi}(T) \,.
\end{equation}
This gives for the baryon chemical potential at the onset of quark saturation 
\begin{equation}
\label{eq:muOSpi}
\muBsat(T)^2 = 2 \left(\frac{\Lambda^2}{\Nc} - \frac{\pi^2}{6}\, T^2\right) + M_N^2 - 2\,\frac{\Lambda^2}{\Nc}\,C_{\pi}(T) \,.
\end{equation}
For fixed $T$, the contribution stemming from the pions reduces the onset value of the baryon chemical potential for quark saturation, see Eq.~\eqref{eq:SF-SaturationCondition}, because $C_{\pi}(T)$ is positive. 
This is shown in Fig.~\ref{fig:HRGSaturation}, where we solve the onset condition in Eq.~\eqref{eq:fQsatBM} numerically for arbitrary $T$. The full result confirms the trend expected from the low-$T$ Sommerfeld expansion. 
In Fig.~\ref{fig:HRGSaturation}, the dashed horizontal line at $T_c=0.158$~GeV marks the deconfinement transition temperature at vanishing baryon chemical potential known from lattice QCD~\cite{Borsanyi:2020fev}. We regard this as a first upper temperature limit in our considerations, above which Quarkyonic Matter cannot be expected. We observe that up to this temperature the onset baryon chemical potential is affected only very little by the pion contributions as one would expect from the $1/N_c^2$ suppression.

We may now extend the discussion of Eq.~\eqref{eq:fQNpi} toward including more hadrons as well as their resonances up to a certain mass cutoff. In a hadron resonance gas (HRG) model, the interactions among hadrons are effectively condensed into the existence of resonances. Each of these particles $i$ has an associated particle chemical potential reading 
\begin{equation}
    \mu_i = B_i\muB + S_i\mu_S + Q_i\mu_Q \, 
\end{equation}
with the quantum numbers $B_i$ for baryon number, $S_i$ for strangeness and $Q_i$ for electric charge and the respective chemical potentials $\muB$, $\mu_S$ and $\mu_Q$. Here, we take again $\mu_Q = \mu_S =0$. While there is already a difference between nucleons and mesons, as discussed for pions above, in their contribution to quark saturation, special care must be taken in the case of strange hadrons. Since we are only discussing light ($u$ and $d$) quark saturation, we can only take into account mesons with $S_i \geq 0$. This implies for example that only $K^+$ and $K^0$ have the relevant quark content and contribute while $K^-$ and ${\bar{K}}^0$ do not. Contributions from strange baryons suffer from a suppression factor of $(\Nc-|S_i|)/\Nc$. Then the quark saturation criterion containing all possible contributions from hadrons and resonances reads
\begin{align}
   1 = & \sum_{i: B_i=1} \frac{d_i}{d_Q} \frac{(\Nc-|S_i|)}{\Nc} \int\, \frac{d^3k}{(2\pi)^3} \,\varphi(k/N_c) \,f^i_{\rm FD}(k) \nonumber\\
   & + \sum_{i: B_i=0, S_i\geq 0} \frac{1}{\Nc}\frac{d_i}{d_Q} \int\, \frac{d^3k}{(2\pi)^3} \,\varphi(k/2) \,f^i_{\rm BE}(k) 
   \label{eq:fQBM} \\ 
   \nonumber
   = & \, \, C_{\rm B}(T,\muB) + C_{\rm M}(T) \,.
\end{align}

In principle, all hadrons and resonances contained in Eq.~\eqref{eq:fQBM} contribute to quark saturation. As all the integrals appearing in Eq.~\eqref{eq:fQBM} are positive, the general effect of each particle contribution is to reduce the baryon chemical potential at the onset of quark saturation for a given temperature. Nonetheless, the most dominant contribution stems from the Delta baryons as is shown in Fig.~\ref{fig:HRGSaturation}. This is because Delta baryons do not suffer from any suppression other than their higher mass compared to nucleons but they profit from a larger degeneracy factor. 
As a consequence, the saturation transition curve becomes strongly bent toward smaller $\muB$ and the transition temperature for a given baryon chemical potential is drastically reduced. Fig.~\ref{fig:HRGSaturation} also shows the impact of all particles included in the sums of Eq.~\eqref{eq:fQBM}. To discuss some systematic error of this result we reduce the considered particle spectrum by demanding a specific mass cut similar to the discussion made in~\cite{Cleymans:2005xv}. We find that on the quantitative level the saturation curve can be shifted by a few percent, see error bars on the HRG model curve, while our qualitative conclusions remain unaffected.

\begin{figure}[t]
      \centering
      \includegraphics[width=0.75\textwidth]{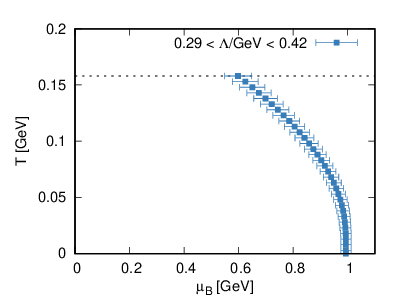}
      \caption{Saturation transition in the $T$-$\muB$ diagram for the HRG model with upper mass limit $2.25$~GeV using $N_c=3$. The error band is determined by varying $\Lambda$ in the interval $1\leq \Lambda/\Lambda_0 \leq 1.44$, where $\Lambda_0\simeq 0.29$~GeV, see also the discussion of Fig.~\ref{fig:nNdensSatT0}.}
      \label{fig:HRGSaturationL}
\end{figure}
A stronger constraint on the location of the saturation curve comes from the uncertainty in the value of the wavefunction momentum $\Lambda$. As we discussed in Fig.~\ref{fig:nNdensSatT0}, an estimate for reasonable parameter values may be obtained by demanding that the onset nucleon density for Quarkyonic Matter formation lies in between one and three times nuclear matter density. For the corresponding range in $\Lambda$-values, $1\leq \Lambda/\Lambda_0 \leq 1.44$ with $\Lambda_0\simeq 0.29$~GeV, the saturation transition curve is depicted in Fig.~\ref{fig:HRGSaturationL}. We regard the error band obtained by the variation of $\Lambda$ as the systematic error in our theory approach to locate the transition curve to Quarkyonic Matter in the QCD phase diagram. While at zero temperature the onset baryon chemical potential may vary by about $38$~MeV, we find at our upper temperature limit a variation of about $100$~MeV in $\muBsat$.

\section{Location in the QCD phase diagram}

It is interesting to discuss the location of the saturation transition curve in the QCD phase diagram, in particular, in comparison with our current knowledge on the latter. The constant upper temperature limit of $T_c=0.158$~GeV we considered so far can only serve as a rough estimate for the maximum temperature at which Quarkyonic Matter may be found at high baryon density. To improve the theoretical estimate for the deconfinement transition line we consider the perturbative QCD expression of the Debye screening mass $m_D$~\cite{LeBellac:1996} which parametrically reads
\begin{equation}
    \label{eq:DebyeScreeningMass}
    m_D^2 (T,\mu) = c_T N_c T^2 + c_\mu \mu^2\,, 
\end{equation}
where $\mu = \muB / N_c$ is the quark chemical potential. We have factored out the dominant $N_c$-dependence in the coefficients $c_T$ and $c_\mu$ so that the remaining constant becomes subleading in $1/N_c$. Assuming that the deconfinement transition occurs when $m_D$ is equal to the QCD scale $\Lambda$, one finds for the baryon chemical potential dependence of the deconfinement temperature 
\begin{equation}
    \label{eq:DeconfinementLinePerturbativeQCD}
    T_c(\mu)^2 = T_c^2 - \frac{c_\mu}{c_T} \frac{\mu^2}{N_c}\,.
\end{equation}
This may serve as an estimate for the deconfinement line in the QCD phase diagram. Quark chemical potential corrections to the horizontal line of constant $T_c$ are suppressed with $1/N_c$ and the deconfinement temperature becomes $\mu$-independent in the large $N_c$ limit. 

\begin{figure}[ht!]
      \centering
      \includegraphics[width=0.75\textwidth]{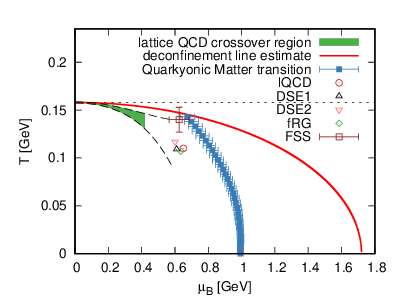}
      \caption{(Color online) Location of the saturation transition curve (represented by full blue squares) as shown in Fig.~\ref{fig:HRGSaturationL} in comparison with the lattice QCD crossover region~\cite{Borsanyi:2020fev} (green band with extended dashed lines) and an estimate for the deconfinement line (red line) based on the Debye screening mass~\cite{LeBellac:1996} in the $T$-$\muB$ diagram. The curve at which quark saturation occurs is now shown only up to the deconfinement temperature at non-zero $\muB$. At this intersection point we find a triple point separating the phases of hadronic matter, deconfined matter and Quarkyonic Matter (located in the region between the curve represented by the full blue squares and the red line). For comparison, we also show current estimates for the location of the chiral critical point based on a continuum estimate of lattice QCD calculations~\cite{Clarke:2024ugt}, the functional methods in~\cite{Gao:2020fbl} (DSE1),~\cite{Gunkel:2021oya} (DSE2) and in~\cite{Fu:2019hdw} (fRG), and the recent finite-size-scaling analysis in~\cite{Sorensen:2024mry}.
      }
      \label{fig:HRGSaturationHRGA}
\end{figure}
Our result for the saturation transition curve as shown in Fig.~\ref{fig:HRGSaturationL}
should now end at the deconfinement line, as depicted for $N_c=3$ in Fig.~\ref{fig:HRGSaturationHRGA}. This happens at a baryon chemical potential of approximately $0.63\leq\muB/$GeV$\leq 0.72$. At the intersection, we find a triple point which separates the phases of hadronic matter, deconfined matter and Quarkyonic Matter. According to these estimates, Quarkyonic Matter can possibly be found in the region between the curve of full blue squares and the red line in Fig.~\ref{fig:HRGSaturationHRGA}. At $T=0$, this corresponds to a range of about $M_N\leq\muB\leq \sqrt{3}M_N$. The found triple point lies roughly in the ballpark of current estimates for the location of the chiral critical point based on lattice QCD~\cite{Clarke:2024ugt}, QCD functional methods~\cite{Gao:2020fbl,Gunkel:2021oya,Fu:2019hdw} and a recent finite-size-scaling analysis~\cite{Sorensen:2024mry}.

For $N_c=3$, the behavior of our estimate for the deconfinement line in Eq.~\eqref{eq:DeconfinementLinePerturbativeQCD} resembles that of the crossover region determined in lattice QCD~\cite{Borsanyi:2020fev}. This is also shown in Fig.~\ref{fig:HRGSaturationHRGA}, where we depict the crossover region of the chiral phase transition as given by a Taylor series up to order $\muB^4$ with the bandwidth determined by the given errors in the expansion parameters~\cite{Borsanyi:2020fev}. This parametrization is shown up to roughly $\muB/T=3$ (green band) but we extend the representation of the upper and lower limits for $T_c(\muB)$ up to $\muB=0.58$~GeV (dashed lines). The lattice QCD crossover band is a bit stronger bent than our estimate for the deconfinement line. Within current errors and based on an extrapolation that is certainly far outside the domain of confidence one may still conclude that one should find a triple point and a Quarkyonic Matter phase at higher $\muB$ in the QCD phase diagram. 

\begin{figure}[t]
      \centering
      \includegraphics[width=0.75\textwidth]{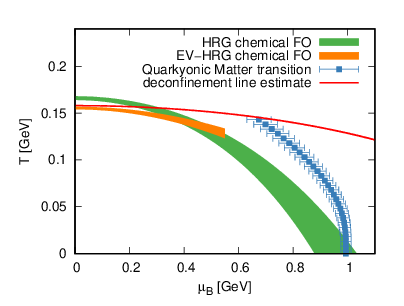}
      \caption{(Color online) Comparison of the phase boundary formed by the deconfinement line discussed in Fig.~\ref{fig:HRGSaturationHRGA} and the quark saturation transition curve presented in Fig.~\ref{fig:HRGSaturationL} with the empirical decoupling regions (chemical freeze-out (FO) curves) determined using a HRG model~\cite{Cleymans:2005xv} (green band) and an excluded volume HRG model~\cite{Alba:2016hwx} (orange band).}
      \label{fig:HRGSaturationHRGB}
\end{figure}
The existence of a Quarkyonic Matter phase could have an impact on physical observables, notably in low energy heavy-ion collisions and for astrophysical observations. It is often advocated~\cite{Andronic:2017pug,Braun-Munzinger:2014lba,Braun-Munzinger:2018yru}, that the measurement of the chemical freeze-out curve, at which the chemical composition of the matter produced in a heavy-ion collision is essentially fixed, helps delineating the QCD phase diagram. Assuming that the decoupling temperatures supposedly measured in the experiments are anywhere close to the transition to Quarkyonic Matter, one might hope that the characteristics of Quarkyonic Matter, for example its different squared speed of sound compared to ordinary baryonic matter~\cite{McLerran:2018hbz,Fujimoto:2023mzy}, affect the experimental data. Such a comparison is done in Fig.~\ref{fig:HRGSaturationHRGB} showing the boundary marked by the deconfinement line and the Quarkyonic Matter transition in contrast to parametrizations of the chemical freeze-out curves based on a HRG model~\cite{Cleymans:2005xv} and an excluded volume HRG model~\cite{Alba:2016hwx}. While the former includes experiments down to collision energies per nucleon of $0.8$~GeV~\cite{Cleymans:2005xv}, the latter stops at a collision energy per nucleon of $20$~GeV~\cite{Alba:2016hwx}. 

In heavy-ion collisions at lowest energies, a substantial fraction of the system passes through the region of high baryon densities $\geq n_0$, where our estimate for Quarkyonic Matter falls (see, e.g.~\cite{Bravina:2008ra, Oliinychenko:2022uvy, Taya:2024zpv}). It is an interesting task for future work to identify potential observables which are sensitive to the interesting phase space structure of baryons and quarks in Quarkyonic Matter (see Ref.~\cite{Torrieri:2013mk} for an earlier attempt). Comparing with the freeze-out curve in Fig.~\ref{fig:HRGSaturationHRGB}, these observables need to be robust enough to survive the evolution from the Quarkyonic Matter region to the decoupling of particles from the system. In particular, fixed target experiments should be able to detect the Quarkyonic Matter phase if it imprints on observables. From an astrophysical point of view, Quarkyonic Matter is expected to have an impact on observations, notably those related to neutron star properties and the dynamical evolution of (binary) neutron star systems.

\section{Summary and Conclusion}

The onset of Quarkyonic Matter coincides with the saturation of the phase-space density of quarks.
Its location in the $T-\muB$ plane of the QCD phase diagram can be determined without the knowledge of the dynamic properties of Quarkyonic Matter once the relation between the phase-space densities of quarks and hadrons is specified. The boundary between ordinary nuclear matter and Quarkyonic Matter can then be identified as the breakdown of the ideal gas or hadronic gas description of the matter when the density is increased. We thus determine the quark saturation curve as the line where the phase-space density of quarks becomes one and thus saturates, i.e.~no more quarks can be added to this phase space region due to the Pauli exclusion principle. In this work, we located semi-quantitatively the quark saturation curve in the QCD phase diagram including a systematic analysis regarding the uncertainty in the location of the curve in the $T$-$\muB$ plane. The sources of uncertainty we considered are the model parameters which set the confinement scale $\Lambda$ in the theory and the hadron species that contribute to the phase-space density of quarks. We constrained the range of $\Lambda$ from the saturation properties at $T=0$. We vary it between $0.29\,\text{GeV}$ and $0.42\,\text{GeV}$ so that the quark saturation takes place between one to three times the nuclear matter density. Another source of uncertainty is the probability distribution of quarks inside a hadron, for which we studied here the Yukawa approach previously applied in the Ideal Dual Quarkyonic (IdylliQ) model. It remains an interesting question if there is some deeper justification for this ansatz, or if experimentally motivated input would be better suited. 

Qualitatively, we see that the quark saturation curve is similar to the expected chiral transition line of QCD. Increasing the temperature, the onset of Quarkyonic Matter occurs at lower baryon chemical potential. In our quantitative study we include subsequently nucleons, pions, Delta baryons and finally all hadrons and their resonances up to a mass cut of $2.25$~GeV. The meson contribution is suppressed relative to that of the baryons by $1/\Nc^2$ and thus mesons have only a small impact on the curve up to the deconfinement temperature. The contribution of strange baryons is suppressed due to their reduced light quark content. As a consequence we identify the Delta baryon to have the strongest single particle impact on the bending of the quark saturation curve. Nonetheless, for all particles, their inclusion reduces the saturation temperature at fixed baryon chemical potential. 

Quarkyonic Matter is not supposed to exist in the deconfined phase. We could therefore constrain the region of Quarkyonic Matter in the QCD phase diagram by restricting it with the deconfinement line at high temperatures and densities. At large $\Nc$, deconfinement is independent of $\muB$ and we put it at the pseudocritical temperature $T_c$ obtained from lattice QCD at zero baryon chemical potential. In order to include finite $\muB$ effects in the finite-$\Nc$ world, we estimate confinement to occur where the Debye screening mass equals the scale $\Lambda$ of QCD. We thus find a triple point where QGP, hadrons and Quarkyonic matter meet in the QCD phase diagram. This region is located consistently at higher temperature and $\muB$ than the decoupling curve of particle species in heavy-ion collisions, but within the reach of low-energy heavy-ion collisions. 

Future work needs to set up a consistent thermodynamic description of Quarkyonic Matter at finite temperature as well as to identify promising observables of Quarkyonic Matter in heavy-ion collisions. 

\begin{acknowledgments}
    M.B. and M.N. thank the Institute for Nuclear Theory at the University of Washington for its kind hospitality and stimulating research environment.
    This research was supported in part by the INT's U.S. Department of Energy grant No. DE-FG02-00ER41132.
    M.N. acknowledges support from the long term visitor program at the INT.
\end{acknowledgments}

\appendix
\section{Sommerfeld expansion for the saturation condition equation\label{sec:A1}}

We may formulate the onset condition for quark saturation Eq.~\eqref{eq:cond_onset} as an integral over the nucleon energy as 
\begin{equation}
    \frac{\Lambda^2}{N_c} = \int_{M_N}^\infty dE_N E_N \,e^{-\sqrt{E_N^2-M_N^2}/(N_c \Lambda)} f^N_{\rm FD}(E_N) \,.
    \label{eq:A1-cond_onset}
\end{equation}
From this equation one obtains the necessary baryon chemical potential for the onset of quark saturation at a given temperature. For small $T$, the behavior of $\muBsat(T)$ can be determined by means of a Sommerfeld expansion. In principle, we can perform this expansion for the integral in Eq.~\eqref{eq:A1-cond_onset} but present here the calculation in the large $N_c$ limit as this is sufficient for our reasoning in Sec.~\ref{sec:2}. In this limit, one can approximate the momentum distribution of quarks in Eq.~\eqref{eq:cond_onset} by $\varphi(k/N_c) \simeq 2 \pi^2 N_c /(\Lambda^2 k)$. Simplifying the somewhat heavy notation in Eq.~\eqref{eq:A1-cond_onset}, we write
\begin{equation}
    \frac{\Lambda^2}{N_c} = I_{\rm sat} = \int_{M}^\infty dE \,E \,f(E) \,.
    \label{eq:A1-cond_onset_simplified}
\end{equation}
Introducing the function $h_{\rm sat}(E)$ as 
\begin{equation}
    h_{\rm sat}(E) = \begin{cases}
        E &,\, E\geq M\\
        0 &,\, {\rm else}
    \end{cases}
\end{equation}
one finds 
\begin{align}
    I_{\rm sat} & = \int_{-\infty}^\infty dE\, h_{\rm sat}(E)\, f(E) \\
    & = - \int_{-\infty}^\infty dE\, H_{\rm sat}(E)\, f'(E) \,,
\end{align}
where $f'(E) = - f(E)(1-f(E))/T$ is the derivative of the Fermi-Dirac distribution with
\begin{equation}
    f(E) = \frac{1}{1+e^{(E-\mu)/T}}
    \label{eq:A1-fN}
\end{equation}
and we defined 
\begin{equation}
    H_{\rm sat}(E) = \int_{-\infty}^{E} h_{\rm sat}(u)\, du\,.
    \label{eq:A1-Hfunction}
\end{equation}
Here we made use of the fact that the function $h_{\rm sat}(E)$ vanishes as $E\to -\infty$ and that it does not diverge more rapidly than some power of $E$ in the limit $E\to +\infty$. 

The Sommerfeld expansion is an expansion of integrals of the type $I_{\rm sat}$ for small, fixed $T$. Assuming that functions of the type of $H_{\rm sat}$ do not vary too rapidly around $\mu$ one may, without loss of generality, expand $H_{\rm sat}$ in a Taylor series as 
\begin{equation}
    H_{\rm sat}(E) = H_{\rm sat}(\mu) + (E-\mu) H_{\rm sat}'(\mu) + \frac12 (E-\mu)^2 H_{\rm sat}''(\mu)+\dots
\end{equation}
to obtain
\begin{align}
\nonumber
   I_{\rm sat} = & - H_{\rm sat}(\mu) \int_{-\infty}^\infty dE \,f'(E) - H_{\rm sat}'(\mu) \int_{-\infty}^\infty dE \,(E-\mu) \,f'(E) \\
\label{equ:A1-IsatTaylor}
   & - \frac12 H_{\rm sat}''(\mu) \int_{-\infty}^\infty dE \,(E-\mu)^2 \,f'(E) + \dots\,,
\end{align}
where the $\mu$ solving Eq.~\eqref{eq:A1-cond_onset_simplified} corresponds to $\muBsat$ since we assume $\mu_Q=\mu_S=0$. 
The above integrals can be evaluated explicitly. With the variable change $x=(E-\mu)/T$ we find
\begin{align}
    & \int_{-\infty}^\infty dE \, f'(E) = -1 \,,\\
    & \int_{-\infty}^\infty dE \, (E-\mu) \,f'(E) = 0 \,,\\
    & \int_{-\infty}^\infty dE \, (E-\mu)^2 \,f'(E) = -\frac{1}{\beta^2} \int_{-\infty}^\infty dx \frac{x^2e^x}{(1+e^x)^2} = -\frac{\pi^2}{3\beta^2} \,,
\end{align}
where $\beta=1/T$. From this, one finally finds as condition for quark saturation 
\begin{equation}
    1 = \frac{N_c}{\Lambda^2} \left(H_{\rm sat}(\muBsat)+\frac{\pi^2}{6}T^2 H_{\rm sat}''(\muBsat)+\dots\right) \,.
    \label{eq:A1-SommerfeldCondition}
\end{equation}
Computing the values for $H_{\rm sat}$ and its second order derivative $H_{\rm sat}''$ at $\muBsat$ explicitly one obtains up to order $T^2$
\begin{equation}
    1 = \frac{N_c}{\Lambda^2} \left(\frac12\left(\muBsat^2-M^2\right)+\frac{\pi^2}{6}T^2 \right) \,.
    \label{eq:A1-SommerfeldCondition-explicit}
\end{equation}
We note that in the large $N_c$ limit all higher-order terms of the Sommerfeld expansion vanish because all derivatives of $H_{\rm sat}$ of order $n\geq 3$ vanish. Solving Eq.~\eqref{eq:A1-SommerfeldCondition-explicit} for $\muBsat$ then gives Eq.~\eqref{eq:SF-SaturationCondition}.

\bibliography{biblio.bib}

\end{document}